\documentclass[%
 reprint,
 superscriptaddress,
 amsmath,amssymb,
 aps,
]{revtex4-1}

\usepackage{graphicx}
\usepackage{dcolumn}
\usepackage{amsthm}
\usepackage{bm}
\usepackage{xcolor}
\newcommand{\ket}[1]{|#1\rangle}
\newcommand{\bra}[1]{\langle#1|}
\newcommand{\stt}[1]{|#1\rangle\langle#1|}
\newcommand{\Tr}[1]{\textrm{Tr}\big[#1\big]}
\newcommand{\sttn}[2]{|#1\rangle\langle#2|}
\newcommand{\iprod}[2]{\langle#1|#2\rangle}
\newtheorem{lemma}{Lemma}

\begin{document}

\title{Approximate quantum state reconstruction without a quantum channel}

\author{Zichen Yang}
\affiliation{School of Physics, Peking University, Beijing 100871, China}

\author{Ze-Yang Fan}%
\affiliation{School of Astronautics, Harbin Institute of Technology, Harbin 150001, China}


\author{Liang-Zhu Mu}
\email{muliangzhu@pku.edu.cn}
\affiliation{School of Physics, Peking University, Beijing 100871, China}

\author{Heng Fan}
\email{hfan@iphy.ac.cn}
\affiliation{Institute of
Physics, Chinese Academy of Sciences, Beijing 100190, China}
\affiliation{CAS Center of Excellence in Topological Quantum Computation, University of Chinese Academy of Sciences, Beijing 100190, China}
\affiliation{Songshan Lake Material Laboratory, Dongguan 523000, China}





\begin{abstract}
We investigate the optimal quantum state reconstruction from cloud to many spatially separated users by
measure-broadcast-prepare scheme without the availability of quantum channel.
The quantum state equally distributed from cloud to arbitrary number of users is
generated at each port by ensemble of known quantum states with assistance of
classical information of measurement outcomes by broadcasting.
The obtained quantum state for each user is optimal in the sense that the fidelity universally achieves the upper bound.
We present the universal quantum state distribution by providing physical realizable measurement bases in the cloud
as well as the reconstruction method for each user. The quantum state reconstruction scheme works
for arbitrary many identical pure input states in general dimensional system.
\end{abstract}

\pacs{Valid PACS appear here}
\maketitle


\section{INTRODUCTION}
In protocols of quantum information processing, entanglement and quantum channel
are in general assumed to be available. However, in a certain
scenario we may need to distribute quantum state
to arbitrary number of users, who are spatially separated, while neither entanglement nor quantum
channel is available.  Each user may prepare their own quantum state according
to classical information broadcasted from the ``cloud'' who can perform measurement
on quantum states need to be distributed. This protocol can be named as classical quantum
state reconstruction (CQSR).
It is known that there is no-cloning theorem for quantum information which
states that an arbitrary quantum state cannot be cloned perfectly \cite{no-cloning,no-cloning-Dieks,HF-report,RMP-cloningReview}.
For spatially separated users, approximate copies of a quantum state can also be
obtained for a number of users by the combination of the quantum cloning machine and
teleportation \cite{teleportation} which needs the resource of entangled states and classical communication \cite{teleclone,YLZhang,HF-report},
differing from CQSR.
One may notice that CQSR can be achieved with the help of
quantum estimation by a measure-and-prepare scheme \cite{Derka-Buzek-Ekert,Bruss-Ekert-Macchiavello,phase-estimation},
but with additional condition that the prepared states should not be entangled \cite{Chiribella1}.
Additionally, CQSR should be physically realizable, which means that the number of measurements
should be finite. We remark that the identically prepared quantum states can be compressed
\cite{ChiribellaPRL2016,ChiribellaPRL20162,ChiribellaFoP}, resulting in that those states may be
broadcasted economically. The quantum broadcast channels are also investigated in Ref.\cite{Wilde}.

The general scheme of CQSR can be shown as in FIG. 1. The cloud will use universal measurement scheme
for arbitrary input states, and broadcast the results of measurement. Each user can prepare the quantum state
by using ensemble of known quantum states agreed in advance,
which are thus in product forms, with probabilities depending on classical information.
Next we generally equate the state estimation with CQSR, but bear in mind the difference
that each user will prepare their state without the assistance of entanglement.
The well-known estimation of quantum states  shows that the {\it mean} fidelity
for input states which are randomly and isotropically distributed can achieve the upper bound \cite{Derka-Buzek-Ekert}.
Here, we focus on the case of {\it universality}
in the sense that each arbitrarily given input can be optimally distributed with the same fidelity.

\section{STATEMENT OF THE PROBLEM}
Here we consider the following case: the arbitrary $M$-copy quantum state $\rho$ in the cloud is to be distributed to $N$ users. Though it will be seen later that $M$ does not necessarily equal to N, we will still start with the $M=N$ case, which meets the requirement of a standard quantum estimation problem.
We first assume that the input is $M$ independent and identically prepared arbitrary pure states in general $d$-dimension Hilbert space $\mathcal{H}$,
$\rho =|\psi \rangle \langle \psi |^{\otimes M}$.  It is known that this state is in the symmetric subspace $\mathcal{H}^M_+$
of $\mathcal{H}^{\otimes M}$ and has a dimension $d_M^+=C_{M+d-1}^{M}$, where $C_{M+d-1}^{M}=\frac {(M+d-1)!}{M!(d-1)!}$.
The basis of symmetric subspace $\mathcal{H}^M_+$ can be denoted by $d$-dimension vectors $\vec{m}=(m_0,m_1\cdots m_{d-1})$ satisfying $\sum_{i=0}^{d-1}m_i=M$,
where $\ket{\vec{m}}$ refers to the symmetric state in which there are $m_i$ copies in the state $\ket{i}$, and $\{\ket{i}\}_{i=0}^{d-1}$ is the
computational basis of Hilbert space $\mathcal{H}$.

The standard quantum estimation process can be considered as a quantum channel $\mathcal{E}(\rho)$ which maps $\mathcal{H}^M_+$ to itself,
\begin{eqnarray}
\tilde{\rho}=\mathcal{E}(\rho)=\sum_{r=1}^R\textrm{Tr}[\hat{O}_r\rho]\stt{\Phi_r}
\end{eqnarray}
where $\hat{O}_r$ is a set of positive operator valued measurement(POVM), and $\stt{\Phi_r}$ is the corresponding guess
in reconstructing the estimated state and also lies in $\mathcal{H}^M_+$.
The case $R$ being finite means physical realizable since measure and broadcast can be implemented finitely.
The completeness requires
\begin{eqnarray}
\sum_{r=1}^{R}\hat{O}_r=\mathbb{I}_+^M, \label{eqn:completenessRelation}
\end{eqnarray}
$\mathbb{I}_+^M$ is the identity of symmetric subspace $\mathcal{H}^M_+$, to make sure the estimation is trace preserving.

\begin{figure}
  \centering
  \includegraphics[width=0.4\textwidth]{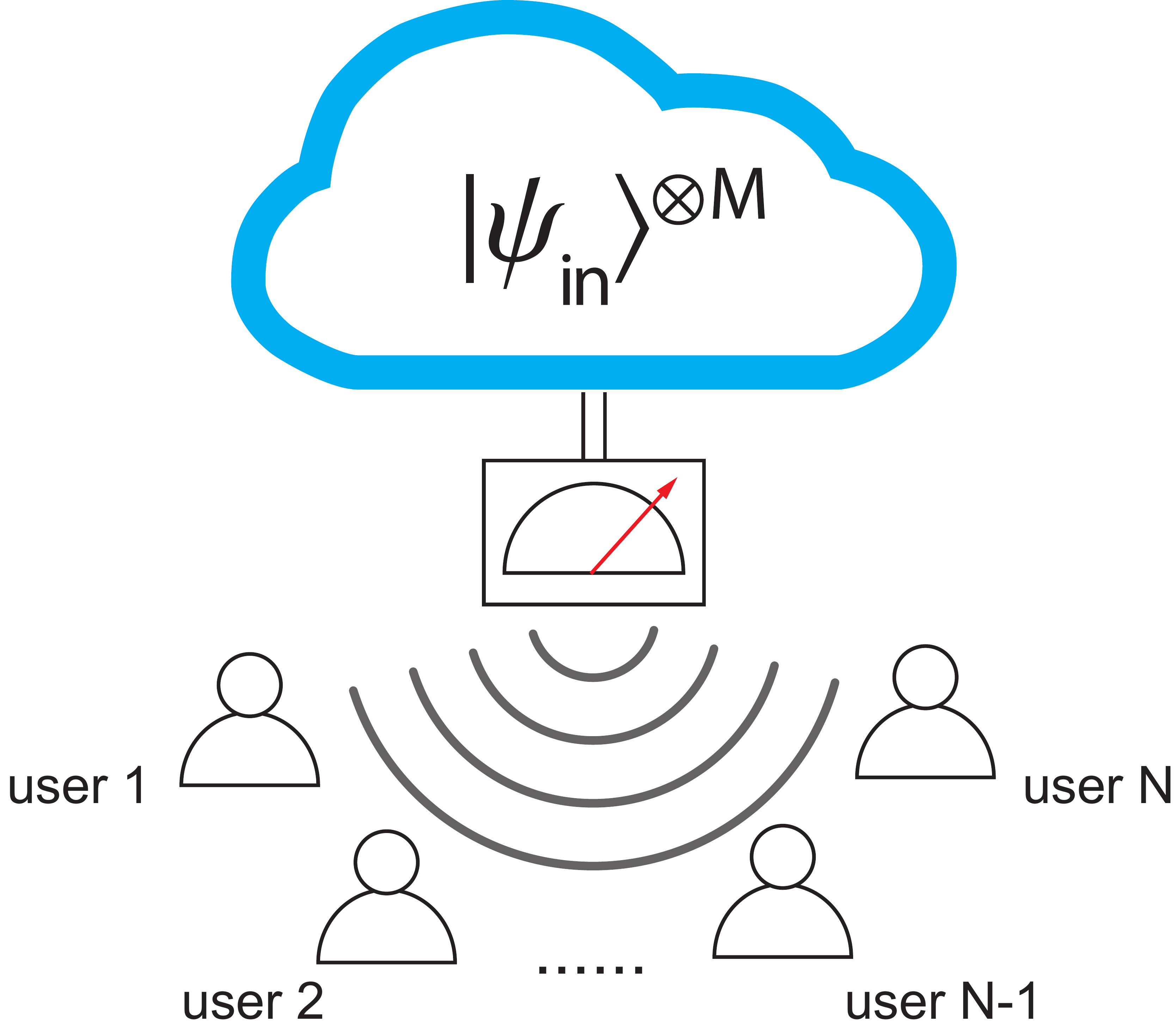}\\
  \caption{Demonstration of quantum state reconstruction in the absence of quantum channel. The input state $\ket{\psi_{in}^{\otimes M}}$ is stored and measured in the cloud. After the measurement, the corresponding measurement result is broadcast to the users. Having obtained the measurement results,
  each user can re-construct the initial state with the optimal fidelity.}\label{cloud}
\end{figure}

For CQSR protocol, we propose that the POVM is performed in the cloud.
Additionally, we need to release the constraint of $M$ users to arbitrary number $N$ of users,
meaning that there is no restriction on number of audiences.
Based on the measurement result, the users reconstruct the state by using known ensemble of states $\{ |\Phi _r\rangle \}$.
We emphasize that state $|\Phi _r\rangle $ is not necessarily the product state for estimation,
however, for spatially
separated users in CQSR,  $|\Phi _r\rangle $ should be in product form but without diminishing the
fidelity.

For simplicity we use the notation $\rho^{(1)}=Tr_{M-1}[\rho]$ and $\tilde{\rho}^{(1)}=Tr_{M-1}[\rho]$, supposing $M=N$. Note again that here $\tilde{\rho}$ relies on the input $\rho=\stt{\psi}^{\otimes M}$. The figure of merit for CQSR can be quantified by the fidelity between a single
copy of the reconstructed state and a single input state $|\psi \rangle $, $f(\psi)=\textrm{Tr}[\rho^{(1)}(\psi)\tilde{\rho}^{(1)}(\psi)]$.
The {\it mean} fidelity is defined as the following form,
\begin{eqnarray}
\bar{f}&=&\int \textrm{d}\psi\textrm{Tr}[\rho^{(1)}(\psi)\tilde{\rho}^{(1)}(\psi)]\nonumber\\
&=&\int \textrm{d}\psi\sum_{r=1}^{R}\textrm{Tr}[\hat{O}_r\stt{\psi}^{\otimes M}]\times
\nonumber \\
&&\textrm{Tr}[\stt{\psi}\textrm{Tr}_{M-1}(\stt{\Phi_r})]\nonumber\\
&=&\int \textrm{d}\psi \sum_{r=1}^{R}\textrm{Tr}\big[\hat{O}_r(U_{\psi}|0\rangle \langle 0|U_{\psi}^\dag)^{\otimes M}]\times\nonumber\\
&&\textrm{Tr}[U_{\psi}|0\rangle \langle 0|U_{\psi}^\dag\textrm{Tr}_{M-1}(\stt{\Phi_r})],
\end{eqnarray}
where ,$U_{\psi}$ is a unitary operator which transforms $|0\rangle $ to $|\psi \rangle $.

Our consideration is to have a {\it universal} fidelity for arbitrary input state.
It is clear that the optimal {\it universal} fidelity cannot exceed the optimal {\it mean} fidelity.
We will find later that these two fidelities are actually the same,
implying that  the {\it universal} fidelity saturates the upper bound.

Here we define the $M$-copy of ensemble of pure states $\{ |\phi _r\rangle \}_{r=1}^R$ with
corresponding probabilities $\{c_r\}_{r=1}^{R}$ as the completely symmetric set (CSS) if it satisfies the relation,
\begin{equation}
\sum_{r=1}^{R}c_r\stt{\phi_r}^{\otimes M}=\frac {\mathbb{I}_+^{M}}{d_M^+ }.\label{eqn:Mrankdef}
\end{equation}
It means that the CSS corresponds to an identity in symmetric subspace.

We have the following lemma:
\begin{lemma}
If $\{\ket{\phi_r}\}_{r=1}^R$ and $\{c_r\}_{r=1}^R$ is an $M$-copy CSS, then it is also $M-1,M-2,\cdots ,1$-copy CSS.
\end{lemma}
\noindent The proof is straightforward. Taking trace over one Hilbert space denoted as, ${\rm Tr_1}$, on both sides of Eq.(\ref{eqn:Mrankdef}),
we find that,
\begin{eqnarray}
\sum_{r=1}^{R}c_r\stt{\phi_r}^{\otimes M-1}
=\frac {1}{d_M^+ }{\rm Tr_1}{\mathbb{I}_+^{M}}
=\frac {\mathbb{I}_+^{M-1}}{d_{M-1}^+},
\end{eqnarray}
Here we need the relation,
\begin{equation}
\ket{\vec{m}}=\frac{1}{\sqrt{C_M^L}}\sum_{\vec{k}}^{C(\vec{k})=M-L}\prod_{j=0}^{d-1}\sqrt{\frac{m_j!}{(m_j-k_j)!k_j!}}\ket{\vec{m}-\vec{k}}\ket{\vec{k}},
\end{equation}
where we have used the notation $C(\vec{k})=\sum_{i=0}^{d-1} k_i$.
In the same way we have $\{\ket{\phi_r}\}_{r=1}^R$ and $\{c_r\}_{r=1}^R$ is also the $M-2,M-3,\cdots ,1$-copy CSS.

Obviously the basis of $\mathcal{H}$ can form a $1$-copy CSS since $\sum_{i=0}^{d-1}\frac{1}{d}\stt{i}=\mathbb{I}/d$.
It is known that the states isomorphically distributed in $\mathcal{H}$ can form an arbitrary $M$-copy CSS,
which is also related to the symmetric distribution of information channel \cite{Chiribella1}.
This infinite set takes the following form,
\begin{equation}
\int \textrm{d}\phi\stt{\phi}^{\otimes M}=\frac{\mathbb{I}_+^M}{d_M^+},\quad M=1,2,3\cdots
\end{equation}
where the integral is taken over the Haar measurement, $M$ is an arbitrary natural number.
However, we need the number of measurements to be finite such that it is physically realizable.

\section{OPTIMAL ESTIMATION PROTOCOL}
Now, we present our main result.\\
\noindent {\bf Theorem.} {\it For state distribution to achieve
optimal {\it mean} fidelity, the POVM must be the form of a $M$-copy CSS.
Additionally, to make the fidelity identical for an arbitrary input,
this CSS should also be the order of $(M+1)$-copy.}

To study the optimal fidelity, it is useful to introduce the following operator,
\begin{eqnarray}
\hat{\mathcal{F}}=\int \textrm{d}\psi\stt{\psi}^{\otimes M}\Tr{\stt{\psi}\stt{0}}.
\label{defofF}
\end{eqnarray}
It is proved that the optimal {\it mean} fidelity $\bar{f}$ is upper bounded by the maximal eigenvalue $\lambda{max}$ of $\hat{\mathcal{F}}$
multiplying the dimension $d$, i.e., $\bar{f}\leqslant d_M^+\lambda_{max}$.
The corresponding POVM has to be $\hat{O}_r=\tilde{c}_rU_r^{\otimes M}\stt{\psi_{max}}U_r^{\dag\otimes M}$,
$\tilde {c}_r$ is the probability and $\stt{\psi_{max}}$ is the eigenstate corresponding to the maximal eigenvalue \cite{Derka-Buzek-Ekert}.

By calculations, for dimension $d$,  we can find that the operator $\hat{\mathcal{F}}$ defined in Eq.(\ref{defofF}) is in the diagonal form,
\begin{eqnarray}
\hat{\mathcal{F}}&=&\int d\psi \stt{\psi}^{\otimes M}{\rm Tr}[\ket{\psi}\iprod{\psi}{0}\bra{0}]\nonumber\\
&=&\sum_{\vec{m},\vec{n}}^{M}\sttn{\vec{m}}{\vec{n}}\int d\psi\bra{\vec{m}}(\stt{\psi})^{\otimes M}\ket{\vec{n}} {\rm Tr}[\ket{\psi}\iprod{\psi}{0}\bra{0}]\nonumber\\
&=&\sum_{\vec{m},\vec{n}}\sttn{\vec{m}}{\vec{n}}{\rm Tr}[\sttn{\vec{n}}{\vec{m}}\otimes \stt{0}\int d\psi\stt{\psi}^{\otimes M+1}]\nonumber\\
&=&\sum_{\vec{m},\vec{n}}\frac{1}{d_{M+1}^+}\sttn{\vec{m}}{\vec{n}}\sum_{\vec{r}}^{C(\vec{r})=M+1}{\rm Tr}[(\sttn{\vec{n}}{\vec{m}}\otimes \stt{0})\stt{\vec{r}}]\nonumber\\
&=&\sum_{\vec{m}}^{M}\frac{1}{d_{M+1}^+}\sttn{\vec{m}}{\vec{m}}\frac{m_0+1}{M+1},
\end{eqnarray}
where summation is taken over all the basis in $\mathcal{H}_+^M$.
For this diagonal matrix $\hat{\mathcal{F}}$, the largest eigenvalue corresponds condition $m_0=M$,
$\lambda_{max}=d_{M+1}^+$, the corresponding eigenstate is $\ket{0}^{\otimes M}$.
The POVM thus takes the form of $M$-identical copies, $\hat{O}_r=\tilde{c}_r(U_r\stt{0}U_r^{\dag})^{\otimes M}=c_rd_M^+\stt{\phi_r}^{\otimes M}$, where $\ket{\phi_r}=U_r\ket{0}$.
 On the other hand, the completeness relation (\ref{eqn:completenessRelation}) requires $\{\ket{\phi_r}\}_{r=1}^R$ and $\{c_r\}_{r=1}^R$ to be a $M$-copy CCS.
The optimal fidelity of state estimation is $\bar{f}_{opt}=\frac{M+1}{M+d}$.
This fidelity is the same as the optimal fidelity of a $M\to\infty$ quantum cloning machine.
The relationship between the fidelity of state estimation and that of the cloning machine
is already known, see \cite{HF-report} and the references therein.
Here we specifically point out that the POVM takes the form as $\hat{O}_r=c_r(U_r\stt{0}U_r^{\dag})^{\otimes M}$,
which simplifies the original result $\hat{O}_r=c_rU_r^{\otimes M}\stt{\psi_{max}}U_r^{\dag\otimes M}$,
where $|\psi _{\max}\rangle $ is generally unknown and may not necessarily be a product state \cite{Derka-Buzek-Ekert}.

However, even if the state estimation achieves optimal {\it mean} fidelity, it is still far from enough,
because for some input states, the fidelity could be undesirably small, which is an unwanted case.
Here we further demand that CQSR yields the {\it universal} fidelity for any input state.
Obviously the {\it universal} fidelity is upper bounded by the
{\it mean} fidelity, namely $\frac{M+1}{M+d}$.
We now prove that this upper bound is achievable for a $(M+1)$-copy CSS.

We can consider the input to be $M$-copy pure states $|\psi \rangle ^{\otimes M}$,
which is in the symmetric subspace. Here, we present a more general form for an arbitrary
matrix in the symmetric subspace for the input,
\begin{eqnarray}
\rho=\sum_{\vec{m},\vec{n}}A_{\vec{m},\vec{n}}\ket{\vec{m}}\bra{\vec{n}}.
\end{eqnarray}
Simply, we know that $\stt{\psi}^{\otimes M}\in \rho $, meaning that the form of identical pure states
is a special case.
After tracing out $M-1$ copies, the single copy state is
\begin{eqnarray}
\rho^{(1)}=\frac{1}{M}\sum_{\alpha,\beta=1}^d\sum_{\vec{m},\vec{n}}A_{\vec{m}\vec{n}}\sqrt{m_\alpha n_\beta}\ket{\alpha}\bra{\beta}\delta_{\vec{m}-\vec{\alpha},\vec{n}-\vec{\beta}},
\end{eqnarray}
here $\vec{\alpha}$ denotes the vector with its $\alpha$-th entry to be $1$ and other entries to be $0$.
If the POVM is $(M+1)$-copy CSS, $d_{M+1}^+\sum_{r=1}^{R}c_r\stt{\phi_r}^{\otimes M+1}=\mathbb{I}_+^{M+1}$,
after some calculations, we can find that the single copy of the output state takes the form,
\begin{eqnarray}
\tilde{\rho}^{(1)}&=&{\rm Tr}_{M-1}[\mathcal{E}(\rho)]\nonumber\\
&=&\sum_{\vec{m},\vec{n}}A_{\vec{m}\vec{n}}\sum_{r=1}^{R}d_M^+ c_r{\rm Tr}\big[\stt{\phi_r}^{\otimes M}\sttn{\vec{m}}{\vec{n}}\big]\stt{\phi_r}
\nonumber \\
&=&\frac{M}{M+d}\rho^{(1)}+\frac{1}{M+d}\mathbb{I}.
\end{eqnarray}
The calculation details can be found in appendix.
These results show that in the sense of single copy state,
the CQSR is equivalent to a polarization channel with a {\it universal} fidelity $F=\frac{M+1}{M+d}$.
So the single copy output state is written universally as the input state with a shrinking factor and a completely mixed state with
a corresponding probability.
For identical pure input states $\rho=
\stt{\psi}^{\otimes M}$, we have $\rho ^{(1)}=\stt{\psi }$.
We emphasize that the fidelity is defined between single input and output states.

Here we would like to address more discussions upon the number of users, $N$ and the copy number of input state, $M$. The general quantum estimation scheme requires the input and output state to have the same copy number. However, as we used single copy fidelity $\textrm{Tr}[\rho^{(1)}\tilde{\rho}^{(1)}]$ instead of overall fidelity $\textrm{Tr}[\rho\tilde{\rho}]$ as the figure of merit, the preparation state may take the direct product form. During this process we actually discarded all the entanglement contained in the original state, which enables us to go beyond the quantum estimation scheme to extend the user number to arbitrary $N$. Correspondingly the definition of single copy fidelity is slightly modified from $\textrm{Tr}_{M-1}[\cdot]$ to $\textrm{Tr}_{N-1}[\cdot]$ for the partial trace, while leaving the main conclusions of this paper unchanged.

For the protocol of CQSR, the importance of
our results is that we only need to find a $(M+1)$-copy CSS, state $\stt{\psi} ^{\otimes M}$ can
be optimally distributed to arbitrary number of users, provided each user can reconstruct their
quantum state by known ensemble of states based on the classical information
broadcasted. It is then crucial that the CSS contains only finite number of states, so that it is
physically realizable.
Operationally, by using $(M+1)$-copy CSS with finite number of states, we can optimally
distribute quantum state to arbitrary number of spatially separated parties without quantum channel.
We remark that the optimal fidelity corresponds to that of universal quantum cloning machine for
infinite copies, however, the cloning machine needs quantum channel to achieve this aim.

\section{EXAMPLES}
In the following we show the protocol of CQSR by
two insightful examples.

{\it Example A:} First let us consider the case where a single qudit (state in d-dimension Hilbert space)
is measured and broadcasted.
Our results suggest that if a 2-copy CSS with finite states is found,
a single qudit can be distributed with the optimal fidelity $\frac{2}{d+1}$.
To construct this CSS set, we introduce the so-called mutually unbiased bases (MUBs), see for example \cite{Roychowdhury,FanPRL}.
For a Hilbert space with dimension $d$, the MUBs contain $d+1$ sets of orthogonal basis $\{\ket{\psi_{t}^{k}}\},t=0,\dots d-1,k=0,\dots d$.
Any states belong to different basis $\ket{\psi_t^{k}}$ and $\ket{\psi_{t'}^{k'}}
(k\neq k')$ satisfy the condition, $|\langle\psi_{t}^{k}|\psi_{t'}^{k'}\rangle|=1/{\sqrt{d}}$, meaning
unbiased for all states.
The construction of MUBs for the case that $d$ is an odd prime number is already well-studied and known to take the following form,
$\ket{\psi_t^0}=\ket{t},
\ket{\psi_t^k}=\frac{1}{\sqrt{d}}\sum_{j=0}^{d-1}(\omega^t)^{d-j}(\omega^{-k})^{s_j}\ket{j},\quad (k\neq 0),t=0,\dots d-1$,
where $\{\ket{j}\}_{0}^{d-1}$ is the computational basis, $s_j=j+\cdots +(d-1)$ and $\omega=\exp(2\pi i/d)$.

We point out that MUBs set constitutes a 2-copy CSS,
 \begin{eqnarray}
 \frac{1}{d(d+1)}\sum_{k=0}^{d}\sum_{t=0}^{d-1}\ket{\psi_t^k}\bra{\psi_t^k}^{\otimes 2}
 =\frac {\mathbb{I}^2_+}{d_2^+}.
 \end{eqnarray}
This identity can be proved by direct calculations, see appendix. According to our results,
we know that by measurement corresponding to MUBs, a single qudit can be optimally distributed without
the availability of quantum channel,
\begin{eqnarray}
\tilde {\rho }&=&\frac{1}{d(d+1)}\sum_{k=0}^{d}\sum_{t=0}^{d-1}
{\rm Tr}(\ket{\psi_t^k}\bra{\psi_t^k}\rho )\ket{\psi_t^k}\bra{\psi_t^k}
\nonumber \\
&=&\frac {1}{d+1}\rho +\frac {1}{d+1}I_d.
\end{eqnarray}
The fidelity is $F=2/(d+1)$ which is optimal. Explicitly, the state $\rho $ is measured in the cloud
by projective measurement corresponding to MUBs, the results are broadcasted. Based on broadcasting information,
each user can construct a quantum state $\tilde {\rho }$ by ensemble states of MUBs with optimal fidelity.

However, the MUBs set is not a general $(M+1)$-copy CSS for $M\ge 1$. We propose that the construction of general
$(M+1)$-copy CSS should be an open problem.

{\it Example B:} Now we consider the qubit situation for case $M$=2, $d$=2. The $2$-dimension MUBs can also be applied to this problem,
where MUBs correspond to the known $6$ bases denoted as, see for example \cite{HF-report},
\begin{eqnarray}
\ket{0},\quad&\ket{+}=\frac{1}{\sqrt{2}}(\ket{1}+\ket{0}),\quad&\ket{\widetilde{+}}=\frac{1}{\sqrt{2}}(\ket{1}+i\ket{0})\nonumber\\
\ket{1},\quad&\ket{-}=\frac{1}{\sqrt{2}}(\ket{1}-\ket{0}),\quad&\ket{\widetilde{-}}=\frac{1}{\sqrt{2}}(\ket{1}-i\ket{0}).\nonumber
\end{eqnarray}
By straightforward calculation, one can find that the 6 states form a $3$-copy CSS,
\begin{eqnarray}
\frac{1}{6}\sum_{\alpha =0,1,+,-,\widetilde{+},\widetilde{-}}\stt{\alpha }^{\otimes 3}=\frac {\mathbb{I}_+^3}{d_3^+}.
\end{eqnarray}
With these 6 bases, one can estimation two identical qubits $|\psi \rangle ^{\otimes 2}$ with optimal fidelity,
\begin{eqnarray}
\tilde {\rho }=\frac {1}{6}\sum _{\alpha }{\rm Tr}(\stt{\alpha }^{\otimes 2}\stt {\psi }^{\otimes 2})\stt{\alpha }^{\otimes N},
\end{eqnarray}
where we write explicitly $N$ in the equation to point out that the number of users $N$ is arbitrary. One can check that
a single qubit output takes the form,
 \begin{eqnarray}
\tilde {\rho }^{(1)}=\frac {1}{2}\stt{\psi }+\frac {1}{4}I_2.
\end{eqnarray}
The fidelity is optimal corresponding to universal quantum cloning machine $2\rightarrow \infty$, which confirms that
our method is applicable.

We emphasize here that the MUBs-constructed CSS is only valid for limited cases. For arbitrary $M$ and $d$,
the completeness relationship is not fulfilled. On the other hand,
we conjecture that $\infty$-copy CSS could only be realized by infinite sets.
If it is true, then any effort to find out a physical realizable finite CSS would be futile,
making the construction of CSS of arbitrary dimension and copies a crucial task.
However, when given a fixed copy number $M$ and dimension $d$, the construction of $M$-copy CSS could be achievable.
Assume that the POVM $\{\hat{O}_r\}_{r=1}^R$, or more specifically, the states $\stt{\phi_r}$, are randomly given,
then one only need to find out a set of positive numbers $\{c_r\}_{r=1}^R$ to satisfy the completeness relationship (\ref{eqn:Mrankdef}).
This simplifies the CSS construction to solving $d_M^{+}(d_M^{+}+1)/2$ linear equations with $R$ unknown variables. By increasing $R$,
which is the total number of POVMs contained in CSS, these equations will be heavily under-determined so that there are enough free parameters to make the $R$ unknown variables all positive.
However it remains a complicated task when $M$ is very large and decreasing the number of equations should be considered.
It is proved in \cite{Derka-Buzek-Ekert} that by applying a set of  rotations $\ket{\phi_r^m}=\exp{(i\tilde{X}\theta_m)}\ket{\phi_r}$, where operator $\tilde{X}$ and constant $\theta_m$ are carefully chosen, one can decrease the number of equations to $d_M^+$, that is, as long as the diagonal elements in ($\ref{eqn:completenessRelation}$) is satisfied, the off-diagonal elements are satisfied as well.

\section{Conclusion}
In conclusion, we have studied the CQSR protocol meaning quantum state reconstruction method in the absence of quantum channel and
provided a physical realizable measurement-and-prepare scheme which achieves the optimal mean fidelity.
The measurement bases of an optimal CQSR must take the form of $M$-copy CSS.
The universal case is also taken into consideration, and we prove that to make the fidelity uniform for arbitrary input,
one only needs to further require the bases to be $(M+1)$-copy CSS.
Two examples for qudit and qubit are given to show the applicable of our method.
We expect that CQSR may stimulate new attention in studying quantum information distribution and concentration.

\section{APPENDIX}
\emph{Single copy outcome in quantum state reconstruction.}---
After the quantum state distribution process, the single copy state of the outcome $\tilde{\rho}$ is,
\begin{eqnarray}
\tilde{\rho}^{(1)}&=&{\rm Tr}_{M-1}[\mathcal{E}(\rho)]\nonumber\\
&=&\sum_{\vec{m},\vec{n}}A_{\vec{m}\vec{n}}\sum_{r=1}^{R}d_M^+ c_r{\rm Tr}\big[\stt{\phi_r}^{\otimes M}\sttn{\vec{m}}{\vec{n}}\big]\stt{\phi_r}\nonumber\\
&=&\sum_{\vec{m},\vec{n}}A_{\vec{m}\vec{n}}\sum_{r=1}^{R}d_M^+ c_r\sum_{\alpha,\beta=0}^{d-1}\sttn{\alpha}{\beta}\nonumber\\
&&{\rm Tr}\big[\stt{\phi_r}^{\otimes M}\sttn{\vec{m}}{\vec{n}}\big]{\rm Tr}\big[\stt{\phi_r}\sttn{\beta}{\alpha}\big]\nonumber\\
&=&\sum_{\vec{m},\vec{n}}A_{\vec{m}\vec{n}}\sum_{\alpha,\beta=0}^{d-1}\sttn{\alpha}{\beta}{\rm Tr}\big[(\ket{\vec{m}}\otimes\ket{\beta})(\bra{\vec{n}}\otimes\bra{\alpha})\nonumber\\
&&\sum_{r=1}^{R}d_M^+ c_r\stt{\phi_r}^{\otimes M+1}\big]\nonumber
\end{eqnarray}
Then we take into account the CSS relation (4) in the main text, and note that $\mathbb{I}^{M+1}_+=\sum_{\vec{s}}^{C(\vec{s})=M+1}\stt{\vec{s}}$, we have,
\begin{eqnarray}
\tilde{\rho}^{(1)}&=&\frac{d_M^+}{d_{M+1}^+}\sum_{\vec{m},\vec{n}}A_{\vec{m}\vec{n}}\sum_{\alpha,\beta=0}^{d-1}\sttn{\alpha}{\beta}\times\nonumber\\
&&{\rm Tr}\big[(\ket{\vec{m}}\otimes\ket{\beta})(\bra{\vec{n}}\otimes\bra{\alpha}))\sum_{\vec{s}}^{C(\vec{s})=M+1}\stt{\vec{s}}\big]\nonumber\\
&=&\frac{d_M^+}{d_{M+1}^+}\sum_{\vec{m},\vec{n}}A_{\vec{m}\vec{n}}\sum_{\alpha,\beta=0}^{d-1}\sttn{\alpha}{\beta}\times\nonumber\\
&&\times\sum_{\vec{s}}^{C(\vec{s})=M+1}\frac{\sqrt{m_\beta+1}}{\sqrt{M+1}}\frac{\sqrt{n_\alpha+1}}{\sqrt{M+1}}\delta_{\vec{s},\vec{m}+\vec{\beta}}\delta_{\vec{s},\vec{n}+\vec{\alpha}}\nonumber\\
&=&\sum_{\alpha,\beta=0}^{d-1}\sum_{\vec{m},\vec{n}}\frac{A_{\vec{m}\vec{n}}\delta_{\vec{m}+\vec{\beta},\vec{n}+\vec{\alpha}}\sqrt{(m_\beta+1)( n_\alpha+1)}}{M+d}\ket{\alpha}\bra{\beta}
\nonumber \\
\end{eqnarray}
The Kronecker-$\delta$ requires when $\alpha\neq\beta$, we have $m_\beta+1=n_\beta$ and $n_\alpha+1=m_\alpha$, and when $\alpha=\beta$, we have $\vec{m}=\vec{n}$.
Then the above equation takes a more concise form,
\begin{eqnarray}
\tilde{\rho}^{(1)}&=&\sum_{\alpha=0}^{d-1}\sum_{\vec{m}}\frac{m_\alpha+1}{M+d}A_{\vec{m}\vec{m}}\stt{\alpha}\nonumber\\
&+&\sum_{\alpha\neq\beta}\sum_{\vec{m},\vec{n}}\frac{\sqrt{m_\alpha n_\beta}}{M+d}A_{\vec{m}\vec{n}}\ket{\alpha}\bra{\beta}
\end{eqnarray}
By tedious but straightforward calculations, we obtain equation (12) in the main text,
\begin{eqnarray}
\tilde{\rho}^{(1)}=\frac{M}{M+d}\rho^{(1)}+\frac{1}{M+d}\mathbb{I}.
\end{eqnarray}

\emph{Two-copy CSS for d-dimension case.}---
Next we will prove that,
\begin{eqnarray}
\hat{Q}&=&\frac{1}{d(d+1)}(\sum_j |j\rangle\langle j|^{\otimes 2}+\sum_{k=1}^{d}\sum_{t=0}^{d-1}\stt{\psi_t^{(k)}}^{\otimes 2})
\nonumber \\
&=&\mathbb{I}_+^2/d_M^+.
\end{eqnarray}

\begin{widetext}
In fact, direct calculation gives that:
\begin{eqnarray}
\bra{j_1j_1}\hat{Q}\ket{j_2j_2}&=&\frac{1}{d(d+1)}(\sum_{k=1}^{d}\sum_{t=0}^{d-1}\bra{j_1j_1}(\stt{\psi_t^{(k)}})^{\otimes 2}\ket{j_2j_2})\nonumber\\
&=&\frac{1}{d(d+1)}\frac{1}{d}\sum_{k=1}^{d}\sum_{t=0}^{d-1}\omega^{2t(j_2-j_1)+k(s_{j_2}-s_{j_1})}\nonumber\\
\bra{j_1,j_2}\hat{Q}\ket{jj}&=&\frac{1}{d(d+1)}(\sum_{k=1}^{d}\sum_{t=0}^{d-1}\bra{j_1,j_2}(\stt{\psi_t^{(k)}})^{\otimes 2}\ket{jj})\nonumber\\
&=&\frac{1}{d(d+1)}\frac{1}{d^2}\sum_{k=1}^{d}\sum_{t=0}^{d-1}\omega^{t(2j-j_1-j_2)-k(s_{j_1}+s_{j_2}-2s_j)}\nonumber\\
\bra{j_1,j_2}\hat{Q}\ket{j_3,j_4}&=&\frac{1}{d(d+1)}\sum_{k=1}^{d}\sum_{t=0}^{d-1}\bra{j_1,j_2}(\stt{\psi_t^{(k)}})^{\otimes 2}\ket{j_3,j_4}\nonumber\\
&=&\frac{1}{d(d+1)}\frac{\sqrt{2}}{d^2}\times
\sum_{k=1}^{d}\sum_{t=0}^{d-1}\omega^{t(j_3+j_4-j_1-j_2)-k(s_{j_1}+s_{j_2}-s_{j_3}-s_{j_4})}\nonumber
\end{eqnarray}
\end{widetext}
We can verify that only $\bra{jj}\hat{Q}\ket{jj}$ and $\bra{j_1,j_2}\hat{Q}\ket{j_1,j_2}$ type elements are nonzero, which indicates $\hat{Q}$ is diagonalized. Further direct calculation proves that the diagonal elements corresponds to these two types have the same value $2/d(d+1)$, i.e. ,$\hat{Q}=\mathbb{I}_+^2/d_2^+$.\\

\emph{Necessary condition for M-copy universal optimal estimation.}---
Now we prove the necessity for the measurement basis to be (M+1)-copy CSS. Suppose that there exists a set of states which forms an optimal estimation measurement operator,
\begin{eqnarray*}
\sum_{r=1}^R c_r\stt{\psi_r}^{\otimes M+1}=\mathbb{I}_+^{M+1}/d_{M+1}^++\hat{P}
\end{eqnarray*}
Operator $\hat{P}$ lies in symmetric subspace $\mathcal{H}_+^{M+1}$ because the left-hand-side of equation
belongs to the symmetric subspace. We prove that there must be $\hat{P}=0$. The output single copy state is,
\begin{eqnarray}
\tilde{\rho}^{(1)}&=&\frac{M+1}{M+d}\rho^{(1)}+\frac{1}{M+d}\mathbb{I}\nonumber\\
&&+\sum_{k,l=0}^{d-1}\ket{k}\bra{l}{\rm Tr}\bigg [(\rho\otimes \ket{l}\bra{k})\hat{P}\bigg ]
\end{eqnarray}
To makes sure that for arbitrary input the fidelity is optimal, the second term must always equal to 0. that is,
\begin{eqnarray}
\Delta_{lk}={\rm Tr}\bigg [(\rho\otimes \ket{l}\bra{k})\hat{P}\bigg ]=0,\quad \forall \rho\in\mathcal{H}_+^{\otimes M},\ket{k},\ket{l}\in\mathcal{H}\label{eqn:zero}\nonumber
\end{eqnarray}
This condition is satisfied only when $\hat{P}=0$. The following part gives a detailed proof.

Since $\hat{P}\in\mathcal{H}_+^{M+1}$,we apply the following expansion form of the operator:
\begin{equation}
\hat{P}=\sum_{\vec{r},\vec{s}}^{C(\vec{r})=C(\vec{s})=M+1}P_{rs}\ket{\vec{r}}\bra{\vec{s}}
\end{equation}
First consider the diagonal elements $P_{rr}$. Suppose that $r_k\neq0$, choose $\rho=\stt{\vec{r}-\vec{k}}$, (\ref{eqn:zero}) gives that:
\begin{eqnarray}
0=\Delta_{kk}=P_{rr}\times \frac{r_k}{M}\Rightarrow P_{rr}=0
\end{eqnarray}
That is, the diagonal elements are all zeroes.

Then consider the off-diagonal elements $P_{rs}$, suppose $r_k\neq0,s_l\neq0$, and for simplicity, let $\vec{m}=\vec{r}-\vec{k},\vec{n}=\vec{s}-\vec{l}$. For state $\rho=\frac{1}{\lambda_1^2+\lambda_2^2}(\lambda_1\ket{\vec{m}}+\lambda_2e^{i\phi}\ket{\vec{n}})(\lambda_1\bra{\vec{m}}+\lambda_2e^{-i\phi}\bra{\vec{n}})$, where $\lambda_1,\lambda_2,\phi$ are non-negative real numbers, $\phi\in[0,2\pi]$. Then (\ref{eqn:zero}) gives,
\begin{eqnarray}
\Delta_{kl}=\frac{1}{\lambda_1^2+\lambda_2^2}(\lambda_1^2A+\lambda_2^2B+\lambda_1\lambda_2(Ce^{i\phi}+De^{-i\phi}))=0,\nonumber
\end{eqnarray}
which is satisfied for arbitrary $\lambda_1,\lambda_2,\phi$. Here
\begin{eqnarray}
A&=&Tr[(\stt{\vec{m}}\otimes\ket{l}\bra{k})\hat{P}],\\
B&=&Tr[(\stt{\vec{n}}\otimes\ket{l}\bra{k})\hat{P}],\\
C&=&Tr[(\ket{\vec{n}}\bra{\vec{m}}\otimes\ket{l}\bra{k})\hat{P}],\\
D&=&Tr[(\ket{\vec{m}}\bra{\vec{n}}\otimes\ket{l}\bra{k})\hat{P}].
\end{eqnarray}
Then we have $A=B=C=D=0$, and $C=0$ gives,
\begin{eqnarray}
\frac{\sqrt{r_k s_l}}{M+1}P_{rs}=0\Rightarrow P_{rs}=0.
\end{eqnarray}
 That is, the off-diagonal elements are also zeroes. Therefore $\hat{P}=0$, which indicates that the quantum estimation is universal only when its measurement bases are (M+1)-copy CSS.

\emph{Acknowledgements:} This work was supported by the National Key R \& D Plan of China (No. 2016YFA0302104,
No. 2016YFA0300600), the National Natural Science Foundation of China (Nos. 91536108, 11774406),
and Strategic Priority Research Program of Chinese Academy of
Sciences (Grant No. XDB28000000).

\end{document}